\documentclass{IOS-Book-Article}

\usepackage{mathptmx}
\usepackage{soul}\setuldepth{article}
\usepackage[linewidth=1pt]{mdframed}
\usepackage{hyperref}
\usepackage{graphicx}

\def\hb{\hbox to 11.5 cm{}}

\begin{document}

\pagestyle{headings}
\def\thepage{}
\begin{frontmatter}              

\title{Challenges and Considerations in Annotating Legal Data: A Comprehensive Overview}

\markboth{}{November 2023\hb}

\author[A]{\fnms{Harshil} \snm{Darji}\orcid{0000-0002-8055-1376}%
\thanks{Corresponding Author: Harshil Darji, Harshil.Darji@uni-passau.de.}},
\author[A,B]{\fnms{Jelena} \snm{Mitrovic}\orcid{0000-0003-3220-8749}}
and
\author[A]{\fnms{Michael} \snm{Granitzer}\orcid{0000-0003-3566-5507}}

\runningauthor{Darji et al.}
\address[A]{Chair of Data Science, University of Passau, Innstraße 41, 94032 Passau, Germany}
\address[B]{The Institute for Artificial Intelligence Research and Development of Serbia, Serbia}

\begin{abstract}
The process of annotating data within the legal sector is filled with distinct challenges that differ from other fields, primarily due to the inherent complexities of legal language and documentation. The initial task usually involves selecting an appropriate raw dataset that captures the intricate aspects of legal texts. Following this, extracting text becomes a complicated task, as legal documents often have complex structures, footnotes, references, and unique terminology. The importance of data cleaning is magnified in this context, ensuring that redundant information is eliminated while maintaining crucial legal details and context. Creating comprehensive yet straightforward annotation guidelines is imperative, as these guidelines serve as the road map for maintaining uniformity and addressing the subtle nuances of legal terminology. Another critical aspect is the involvement of legal professionals in the annotation process. Their expertise is valuable in ensuring that the data not only remains contextually accurate but also adheres to prevailing legal standards and interpretations. This paper provides an expanded view of these challenges and aims to offer a foundational understanding and guidance for researchers and professionals engaged in legal data annotation projects. In addition, we provide links to our created and fine-tuned datasets and language models. These resources are outcomes of our discussed projects and solutions to challenges faced while working on them.
\end{abstract}


\end{frontmatter}
\markboth{November 2023\hb}{November 2023\hb}

\section{Introduction}

Legal data annotation is a critical step in developing machine learning models and natural language processing tools tailored for the legal domain. The significance of this process stems from the unique nature of legal texts, characterized by their complicated structures, specialized terminology, and the importance of context. While data annotation is a common practice across various fields, the legal sector presents challenges that necessitate a specialized approach.

The initial phase of this process involves selecting a raw data set. Given the multifaceted nature of legal documents, this step is crucial in ensuring that the data captures the complexities inherent to legal language. Once a data set is chosen, the task of text extraction begins. This is a complex endeavor, as legal documents often contain footnotes, references, and other elements that can complicate the extraction process. Moreover, redundant information necessitates rigorous data cleaning to retain essential legal details while eliminating irrelevant content.

Annotation guidelines play a pivotal role in this process. They provide a framework that ensures consistency in annotation, considering the nuances of legal language. Furthermore, the involvement of legal professionals in the annotation process cannot be overstated. Their expertise ensures that the annotated data is not only contextually accurate but also aligns with current legal standards and interpretations.

In addition to these challenges, practical considerations also come into play. For instance, many annotators, especially those from a legal background, may prefer using familiar tools like Microsoft Word for annotation. This preference introduces another layer of complexity, as extracting annotations from resulting formats and converting them into a structured format like CoNLL can be a daunting task. Addressing these practical challenges is essential to ensure the smooth progression of the annotation project and to maintain the quality and integrity of the annotated data. Through this paper, we aim to shed light on these challenges and offer a comprehensive overview to assist researchers and professionals in their legal data annotation projects.

\section{Related work}
Legal data annotation, particularly in the domain of discourse phenomena, has been the subject of extensive research. The intricacies of extracting substantive legal knowledge from legal sources have been highlighted by Santosuosso et al. \cite{r1}, emphasizing the challenges this presents in legal analytics. Their exploration resonates with the complexities faced in the legal data annotation process, emphasizing the need for robust solutions.

Wyner et al. \cite{r2} have explored the challenges associated with the manual extraction of legal rules from legal texts, emphasizing the importance of clear annotation guidelines and the involvement of legal professionals in the annotation process. Their work underscores the complexities of legal language and the challenges of extracting meaningful information from dense legal texts.
The challenges of semantic role labeling in legal texts have been explored by Ceci et al. \cite{r3}, highlighting the unique syntactic and semantic structures present in legal documents. Their work emphasizes the need for specialized tools and approaches tailored to the legal domain.

Rättzén et al. \cite{r4} highlight the challenges in legal data annotation for automated contract review, emphasizing the complexity of translating contract language into machine-readable formats. They note the potential for machine learning to revolutionize contract scrutiny despite the need for meticulous manual annotation and the inherent risks of confidentiality breaches.

The complexities of annotating multiparty discourse in legal texts have been highlighted by Wacholder et al. \cite{r5}, emphasizing the challenges of achieving a reliable Inter-Annotator Agreement (IAA). Their observations resonate with the challenges faced in legal data annotation, where identifying unit boundaries can be remarkably intricate.

Walker et al. \cite{r6} address the need for annotated legal corpora and effective human annotation protocols in legal document analysis. They emphasize the lag in applying advanced analytics to legal texts compared to other fields, attributing this to the absence of practical theories of legal reasoning for software development and the complexity of legal argumentation for AI.

The study by Correia et al. \cite{r7} addresses the challenge of annotating legal documents for Named Entity Recognition (NER) in the context of the Brazilian Supreme Court. It presents a fine-grained legal entity annotation case study, utilizing law students as annotators to create a corpus of 594 decisions.

Urchs et al. \cite{r8} presented a corpus of 32,748 German legal decisions and a subset of 200 randomly chosen judgments annotated for the German legal writing style Urteilsstil. As per the authors, annotating legal texts is a complicated task for machine learning, as it involves defining clear legal terms and including facts under legal definitions. However, legal language is nuanced and context-dependent, making it challenging to annotate legal datasets in detail.

\section{Challenges of Legal Annotations}

Legal data annotation, especially within German legal texts, is filled with complexities that arise from the nature of legal language, the architecture of legal documents, and the subtleties inherent in legal references. In this section, we discuss the similar challenges we faced during our previous works.

\subsection{Dataset structuring and availability}
One of the primary hurdles in the legal NLP domain is the need for well-structured datasets. Existing datasets in the German legal domain often encompass entire legal cases or specific named entities within those cases. However, there is a pressing need for datasets that provide comprehensive texts for each legal reference cited in a legal case, along with specific paragraph texts. This challenge directly impacts the annotation process, as the quality and structure of the dataset determine the accuracy and efficiency of legal data annotation.

One of our previous works explored semantic similarities between legal texts and corresponding referred laws \cite{r9}. This task requires a large number of legal cases as well as legal laws with their definitions. We fulfilled the necessary condition of gathering legal cases by relying on Open Legal Data\footnote{\url{https://de.openlegaldata.io/}} introduced by Ostendorff et al. \cite{r10}

However, all the legal cases in the downloaded raw dataset were in HTML format. Because of this, important information, such as \textbf{\textit{tenor}}, \textbf{\textit{tatbestand}}, \textbf{\textit{gründe}}, and \textbf{\textit{entscheidungsgründe}}, was difficult to extract due to the inconsistency of HTML structure throughout the dataset. We dealt with these inconsistencies by making some assumptions, one of which was to assume that all the important titles, such as \textbf{\textit{tenor}} and \textbf{\textit{gründe}}, are within the \textbf{\textit{h2}} tag. We also made sure the content within \textbf{\textit{h2}} tag is all alphabetic and is within a certain length. The reason is that there were some instances where titles were written as \textbf{\textit{1. tenor}} or \textbf{\textit{t e n o r}}. Considering we were only focusing on a few important titles, it was an excellent choice to make assumptions related to this to avoid any false positives.

After cleaning the raw dataset using such assumptions, we now have a dataset of approximately \textbf{1.1 GBs}, with \textbf{43337} rows and \textbf{12} features (see table \ref{tab:data_example}). This dataset is currently publicly available on Zenodo\footnote{\url{https://zenodo.org/records/6631931}}.

\begin{table}[!h]
	\centering
	\begin{tabular}{|l|c|l|}
	    \hline
		\textbf{Feature} & \textbf{Total} & \textbf{Example content}\\
		\hline
		id & 43337 & 127981\\
		slug & 43337 & ag-volklingen-2002-07-10-5c-c-24102\\
		ecli & 10831 & NaN\\
		date & 43337 & 2002-07-10\\
		court & 43337 & Amtsgericht Völklingen\\
		jurisdiction & 43337 & Ordentliche Gerichtsbarkeit\\
		level\_of\_appeal & 43337 & Amtsgericht\\
		type & 43337 & Urteil\\
		\textbf{tenor} & 36282 & 1. Die Beklagten werden als Gesamtschuldner...\\
		\textbf{tatbestand} & 24243 & Auf die Darstellung des Tatbestandes...\\
		\textbf{gründe} & 27144 & Die Klage ist zulässig und begründet. Die...\\
		\textbf{entscheidungsgründe} & 24038 &  Die Klage ist zulässig und begründet. Di...\\
		\hline
	\end{tabular}
	\caption{Here, \textit{slug} can be used to view the web version of any case by appending it to \textit{https://de.openlegaldata.io/case/[slug]}. ECLI, which stands for European Case Law Identifier, simplifies the accurate citing of rulings from European and national courts.}
	\label{tab:data_example}
\end{table}

\subsection{Information extraction}

As mentioned in the previous section, calculating semantic similarities between legal text and corresponding law also requires access to the definition of the corresponding law. For example, if the following is part of text from a legal case,

\begin{mdframed}
    \textit{Die Klägerin hat ein rechtliches Interesse an der begehrten Feststellung (§ 256 Abs. 1 ZPO), da sie von den ...}
\end{mdframed}

Then, the corresponding law definition will be the law text for the law \textbf{\textit{§ 256 Abs. 1 ZPO}}, which can be found on \textit{Gesetze im Internet}\footnote{\url{https://www.gesetze-im-internet.de/}} as,

\begin{mdframed}
    \textit{Auf Feststellung des Bestehens oder Nichtbestehens eines Rechtsverhältnisses, auf Anerkennung einer Urkunde oder auf Feststellung ihrer Unechtheit kann Klage ...}
\end{mdframed}

As seen in this example, this requires the extraction of similar laws from the available dataset of legal cases. However, the dataset we worked with did not have a section that collected all the laws referred to in every legal case available. This issue can be resolved using regex, which filters and extracts specific data patterns. However, using regex has problems if the given dataset uses different ways for a similar string (e.g., \textit{\textbf{§§} 47 Abs. 1, \textbf{§} 154 Abs. 2, or \textbf{Art.} 6}), as shown in the following examples:

\begin{mdframed}
\begin{enumerate}
\item Let's begin with a simple regex to identify basic references to German legal statutes:
\begin{verbatim}
    §\s*\d+\s*(Abs\.\s*\d+)?\s*(Satz\s*\d+)?
\end{verbatim}
It captures patterns that denote a section (\textbf{§}) followed by a section number and, optionally, subsections and sentences within that section. However, while using this regex on a given set of examples, we ran into the issue of it constantly missing the main section numbers (\textit{like § 433, § 434, etc.}).\\

\item The following regex example is intended to identify specific references to articles in major German legal codes such as the GG, BGB, StGB, and HGB:
\begin{verbatim}
    Art\.\s*\d+\s*(Abs\.\s*\d+)?\s*(Satz\s*\d+)?
    \s*(GG|BGB|StGB|HGB)
\end{verbatim}
However, as with the first regex example, this also failed to identify the main article number in law references. Another issue with this regex is its inability to identify unabbreviated forms of similar German legal codes.\\

\item The syntax of these law references is very inconsistent throughout the dataset, e.g., it often starts with ``\textbf{§}" or ``\textbf{§§}" etc. \cite{r11} An example of regex that can be used to resolve this issue is:
\begin{verbatim}
    (§§?|Art\.)\s*(\d+\w*(?:-\d+\w*)?)
\end{verbatim}
However, this regex is restricted to “\textbf{§}”, “\textbf{§§}”, and “\textbf{Art.}” and cannot be generalized. To make it suitable for similar inconsistencies, it needs to be modified each time one is encountered.
\end{enumerate}
\end{mdframed}

These are just a few examples where regex fails to extract the information due to differences in representation and inconsistent formatting of law references. Regex is precise for well-defined and consistent patterns, but legal documents are not always consistent.

To tackle this problem, we fine-tuned the \textit{bert-base-cased}\footnote{\url{https://huggingface.co/bert-base-cased}} language model using the \textit{Legal-Entity-Recognition}\footnote{\url{https://github.com/elenanereiss/Legal-Entity-Recognition}} dataset. This fine-tuned BERT model is capable of identifying laws in a given legal case with an F1 score of \textbf{99.29} \cite{r12}. Another advantage of fine-tuning a language model over regex is its ability to identify and extract more than just law references. Our fine-tuned BERT model can recognize \textbf{19} different named entities with high precision. We have made it publicly available on HuggingFace\footnote{\url{https://huggingface.co/PaDaS-Lab/gbert-legal-ner}}, which currently has over 3000 downloads.

The mentioned \textbf{19} named entities are:
\textit{Person} (\textbf{PER}), \textit{Judge} (\textbf{RR}), \textit{Lawyer} (\textbf{AN}), \textit{Country} (\textbf{LD}), \textit{City} (\textbf{ST}), \textit{Street} (\textbf{STR}), \textit{Landscape} (\textbf{LDS}), \textit{Organization} (\textbf{ORG}), \textit{Company} (\textbf{UN}), \textit{Institution} (\textbf{INN}), \textit{Court} (\textbf{GRT}), \textit{Brand} (\textbf{MRK}), \textit{Law} (\textbf{GS}), \textit{Ordinance} (\textbf{VO}), \textit{European legal norm} (\textbf{EUN}), \textit{Regulation} (\textbf{VS}), \textit{Contract} (\textbf{VT}), \textit{Court decision} (\textbf{RS}), \textit{Legal literature} (\textbf{LIT}).

\subsection{Manual annotation and expertise}

Building on the previous section, the next part of the process is to create annotations for each extracted law. This includes identifying various parts of a law, such as \textbf{\textit{Artikel}}, \textbf{\textit{Satz}}, \textbf{\textit{Absatz}}, etc. Consider the following example:

\begin{mdframed}
    For a given law \textbf{\textit{§ 14 Abs. 3 Satz 2 SchVG}},
    \begin{itemize}
        \item Artikel = \textit{14}
        \item Absatz = \textit{3}
        \item Satz = \textit{2}
    \end{itemize}
\end{mdframed}

As shown in the above example, some information is easily accessible from a given law. However, other information, such as \textbf{\textit{Gesetzbuch}}, \textbf{\textit{law\_title}}, \textbf{\textit{absatz\_text}} can only be found by visiting the corresponding online link for each law. Such annotations can not be done programmatically and require manual efforts, which is not only labor-intensive but also mandates deep expertise. We tackled this problem with the help of three legal experts who have already passed their first state examination in law. These experts manually annotated extracted law references with following \textbf{21} distinct labels, creating a dataset of \textbf{2944} unique law references:

\begin{mdframed}
{\textit{Gesetzbuch/Norm, Buch , Teil, Abschnitt, Titel, Untertitel, Kapitel, Artikel, Absatz, Buchstabe, Unterabsatz, Satz, Nummer, Buchstabe , Online Link Gesetzbuch, Online Link Exakt, Alternative Schreibeweise 1, Alternative Schreibeweise 2, law\_title, full\_text, absatz\_text}}
\end{mdframed}

The inter-annotator agreement between these three annotators, calculated using Fleiss’ Kappas score, is \textbf{0.87} \cite{r13}. Considering that this dataset, alongside the Open Legal Data, can be used by researchers for various purposes, such as law reference prediction or link prediction, it is now publicly available on HuggingFace\footnote{\url{https://huggingface.co/datasets/PaDaS-Lab/legal-reference-annotations}}.

While on the topic of manual annotation, another challenge we faced during our most recent work was due to the annotation tool being used. Our most recent work focuses on creating a dataset of privacy policies annotated using GDPR-compliant named entities, such as \textit{Data Controller}, \textit{Data Processor}, \textit{Authority}, etc. For this process, we retained two legal experts who annotated law references.

To begin with, we collected privacy policies in English from \textbf{45} different online platforms such as \textit{9gag}, \textit{Amazon}, \textit{Airbnb}, etc. Since the legal experts are familiar with Microsoft Word, these privacy policies were stored in the same format. Both annotators used the comment feature of Microsoft Word to perform annotations using \textbf{33} distinct labels. However, the challenge was extracting the commented parts and comments and storing them in a format suitable for further analysis. Popular libraries such as Aspose\footnote{\url{https://docs.aspose.com/words/python-net/}} and python-docx\footnote{\url{https://python-docx.readthedocs.io/en/latest/}} failed, considering they only managed to extract less than 35\% of total comments.

Finally, we tackled this problem by using \textbf{XPath}, an expression language designed for selecting nodes from an XML document, supporting both queries and transformations. Using Python, when a ``\textbf{\textit{.docx}}" file is treated as a ``\textbf{\textit{.zip}}" file, its output includes ``\textbf{\textit{comments.xml}}" and ``\textbf{\textit{document.xml}}". After extracting comments from ``\textbf{\textit{comments.xml}}" and using it with the following XPath expression on ``\textbf{\textit{document.xml}}", we managed to extract commented parts and corresponding commented labels with utmost precision.

\begin{mdframed}
\begin{verbatim}
# XPath expression to find text associated with the comment
xpath_expr = f'//w:commentRangeStart[@w:id="{k}"]/following:' \
             f':w:t[following::w:commentRangeEnd[@w:id="{k}"]' \
             f' and not(preceding::w:commentRangeStart' \
             f'[@w:id="{int(k)+1}"])]'
\end{verbatim}
\end{mdframed}

This XPath expression can be integrated into a Python script\footnote{\url{https://gist.github.com/harshildarji/8480e6f083021c8bd0fce0d297a9855c}} as it is and can be interpreted as follows:

\begin{mdframed}
$\mathbf{//w:commentRangeStart}$\\
Selects all $<w:commentRangeStart>$ elements in the XML document, regardless of their location.\\\\
$\mathbf{[@w:id=``{k}"]}$\\
Filters the selected $<w:commentRangeStart>$ elements by the $w:id$ attribute value equal to ${k}$. This attribute value will likely be a variable or placeholder being dynamically replaced with an actual value when the XPath expression is used.\\\\
$\mathbf{/following::w:t}$\\
Selects all $<w:t>$ elements that appear after the filtered $<w:commentRangeStart>$ element(\textit{s}).\\\\
$\mathbf{[following::w:commentRangeEnd[@w:id=``{k}"]}$\\
Filters the selected $<w:t>$ elements further by checking if there is a $<w:commentRangeEnd>$ element with a matching $w:id$ attribute value of ${k}$ that appears after the current $<w:t>$ element.\\\\
$\mathbf{and\ not(preceding::w:commentRangeStart[@w:id=``{int(k)+1}"])]}$\\
Additionally filters the selected $<w:t>$ elements by checking if there is no $<w:commentRangeStart>$ element with a $w:id$ attribute value equal to ${int(k)+1}$ that appears before the current $<w:t>$ element.
\end{mdframed}

AAno Considering the fact that this dataset can help improve current privacy policy initiatives, we also plan to make this dataset public in CoNLL format \cite{r14}.

\section{Conclusion}

Efforts to structure and annotate legal datasets, especially in the German legal domain, come with various challenges. The complex nature of legal language and the diverse yet inconsistent structure of legal domains require a careful and detailed approach to dataset creation and annotation. In this paper, we discussed our experiences highlighting the need for a well-structured and consistent dataset. Despite facing challenges with HTML inconsistencies and needing tedious manual annotations, we have created datasets that provide access to clean Open Legal Data and corresponding information for individual laws referred to in legal cases. We also fine-tuned a language model to tackle the problem of extracting information from the available dataset, which allowed for a fine-tuned BERT model capable of recognizing 19 distinct named entities in German legal cases.

We also discussed problems that emerged due to the use of Microsoft Word for annotation purposes and the use of its comments feature to annotate privacy policies with GDPR-compliant named entities. Although using Word as an annotation tool caused concern while extracting the commented parts and corresponding comments/labels, we found a way to do so using XPath expression properly. While investigating the low inter-annotator agreement score, we also inferred that one of the main reasons behind this is the differences in comment boundaries rather than the usual disagreement between annotators.

In this paper, we reintroduced several related datasets and language models we created and fine-tuned, respectively. After solving various challenges during our work on the mentioned projects, we have concluded that it is crucial to understand these challenges as the field of legal data annotation continues to expand. Addressing these challenges ensures the success of the annotation projects and maintains the quality of the annotated data in the legal domain.

\section{Acknowledgements}

\begin{flushleft}
\includegraphics[width=0.25\linewidth]{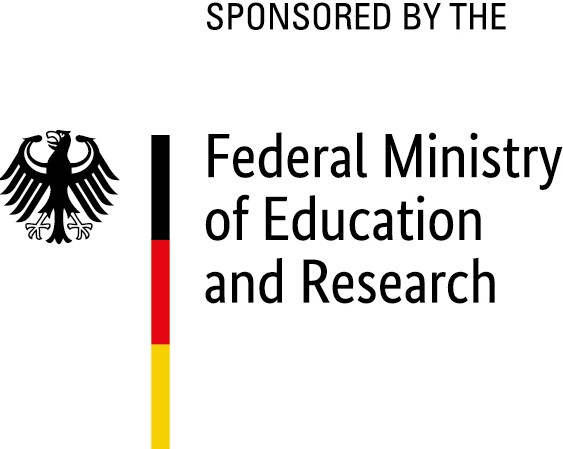}
\end{flushleft} 

The project on which this report is based was funded by the German Federal Ministry of Education and Research (BMBF) under the funding code 01|S20049. The author is responsible for the content of this publication.

\end{document}